# Formation of β-In$_2$Se$_3$ Layers Grown via Selenium Passivation of InP(111)B Substrate

*Kaushini S. Wickramasinghe[a], Candice Forrester[a,b], Martha R. McCartney[c],*

*David J. Smith[c], and Maria C. Tamargo[a,b]\**

[a]Department of Chemistry and Biochemistry, The City College of New York, NY 10031, United States

[b]Chemistry Program, CUNY Graduate Center, New York, NY 10016, United States

[c]Department of Physics, Arizona State University, Tempe, AZ 85287, United States

*Email: mtamargo@ccny.cuny.edu

**ABSTRACT**

Indium selenide, $In_2Se_3$, has recently attracted growing interest due to its novel properties, including room temperature ferroelectricity, outstanding photoresponsivity, and exotic in-plane ferroelectricity, which open up new regimes for next generation electronics. $In_2Se_3$ also provides the important advantage of tuning the electrical properties of ultra-thin layers with an external electrical and magnetic field, making it a potential platform to study novel two-dimensional physics. Yet, $In_2Se_3$ has many different polymorphs, and it has been challenging to synthesize single-phase material, especially using scalable growth methods, as needed for technological applications. In this paper, we use aberration-corrected scanning transmission electron microscopy to characterize the microstructure of twin-free single-phase ultra-thin layers of $\beta$-$In_2Se_3$, prepared by a unique molecular beam epitaxy approach. We emphasize features of the $In_2Se_3$ layer and $In_2Se_3$/InP interface which provide evidence for understanding the growth mechanism of the single-phase $In_2Se_3$. This novel approach for forming high-quality twin-free single phase two-dimensional crystals on InP substrates is likely to be applicable to other technologically important substrates.

Two-dimensional (2D) van der Waals (vdW) materials have garnered much interest over the past decade because of the large variety of 2D compounds having potential applications for next generation electronic and optoelectronic devices[1,2]. The class of 2D vdW materials began to evolve two decades ago, with the discovery of graphene[3,4] which is a semimetal. Since then, the list of 2D materials has been rapidly expanding, and includes insulators such as hexagonal boron nitride (h-BN), transition metal dichalcogenides (TMDCs), such as $MoS_2$, materials from group III-VI family (e.g. $In_2Se_3$, GaSe), and the black phosphorus family (e.g. P, SnS)[5]. Unlike 3D materials, it is easy to engineer the band gap using quantum confinement simply by changing the number of layers in the 2D material[6]. Since these compounds are layered with strong in-plane covalent bonds, while planar layers are held together by weak out-of-plane vdW forces, the layers are easily separable via mechanical exfoliation and can be stacked up with different 2D materials to create novel heterostructures[5]. Either by quantum confinement or by making heterostructures, many interesting physics topics can be studied, for example, superconductivity, magnetism, and charge density waves. Furthermore, the properties the 2D material can be changed by applying an external electrical and magnetic field so that the material can be tuned into a different phase. For example, experiments showed that $TiSe_2$, a semimetal, can be tuned into a superconducting phase by applying an external electric field[7].

Indium selenide, $In_xSe_y$ has recently attracted renewed interest not only due to applications in thermoelectric devices to harness green energy[8] but also owing to some of its novel properties such as room temperature ferroelectricity in the $\alpha$-$In_2Se_3$ phase[9], outstanding photoresponsivity in $\beta$-$In_2Se_3$[10], and exotic in-plane ferroelectricity in $\beta$'-$In_2Se_3$[11,12]. Thus, $In_2Se_3$ has potential applications in energy harvesting, such as solar cells and photodetectors[10], as well as in electronic applications, such as ferroelectric semiconductor field effect transistors[9] and phase change materials for data storage[13]. However, In$_2$Se$_3$ is a complex material with many different polymorphs, known as α, β, β', γ, δ, and κ[14,15]. Hence, it has been challenging to synthesize single-phase In$_2$Se$_3$. There are a few reports of In$_2$Se$_3$ growth using physical vapor transport (PVT)[14] which limits the crystal size, while chemical vapor deposition (CVD)[16] and metalorganic chemical vapor deposition (MOCVD)[17] are employed for scalable synthesis. Yet the overall quality of the material is not good. Among all scalable crystal growth methods, molecular beam epitaxy (MBE) is the preferred technique since it can provide high crystalline quality and controllability of thickness down to the few Ångstrom level. However, there is only one report of In$_2$Se$_3$ grown on sapphire substrates using MBE[18]. Thus, epitaxial growth of scalable In$_2$Se$_3$ is still in its infancy. Hence, it is important to develop a scalable growth method that provides high quality and single-phase In$_2$Se$_3$, especially on substrates that are currently in use for technological applications.

In this study, we have used aberration-corrected scanning transmission electron microscopy (STEM) to conduct an in-depth analysis of twin-free single phase β-In$_2$Se$_3$ as-grown on smooth non-vicinal InP(111)B substrate using a non-conventional MBE growth method developed in our laboratory[19]. Based on this analysis, we also discuss the possible growth mechanism that resulted in the untwinned nature of the material. Understanding and controlling the mechanism at play should enable application of the same approach to growth of other heteroepitaxial structures involving layered, vdW materials on 3D crystalline substrates. It may also enable the possibility of stabilizing the In$_2$Se$_3$ α-phase which is of important technological interest owing to its ferroelectric properties.

Previously, we have presented extensive evidence for the twin-free nature of In$_2$Se$_3$ grown on smooth non-vicinal InP(111)B substrates using high resolution X-ray diffraction (HR-XRD) measurements[19] which was then used as a template to achieve twin-free Bi$_2$Se$_3$. In this paper, we characterize the structural properties of this twin-free Bi$_2$Se$_3$ and In$_2$Se$_3$ using cross-sectional STEM with special emphasis on features of the In$_2$Se$_3$ layer and its interfaces, which provide the evidence needed to understand the growth mechanism of the twin-free In$_2$Se$_3$ layer grown by the unique non-conventional technique[19].

A cross-sectional high-angle annular-dark-field (HAADF) STEM image of the sample, which consists of two distinct layers, Bi$_2$Se$_3$ and In$_2$Se$_3$, with thicknesses of 65 nm and 7 nm, respectively, is shown in Figure 1a. The Bi$_2$Se$_3$ layer displays a defect-free nature without anti-phase domains and no indication of twin boundaries or dislocations. Further cross-sectional HAADF-STEM images taken at widely separated locations on the same sample, confirm the excellent crystal quality (see Supplementary Material). In contrast, reports of twin-free Bi$_2$Se$_3$ grown on rough non-vicinal InP(111)B substrates have previously shown the presence of anti-phase domains due to variations in substrate height[20] and Bi$_2$Se$_3$ grown on flat InP(111)B has shown twin boundaries as well as dislocations[21]. Figure 1b shows well-ordered Bi$_2$Se$_3$ quintuple layers with an overlay of an atomistic model (Figure 1c), which confirms the presence of the rhombohedral crystal structure with the space group $R\bar{3}m$. Additional HAADF-STEM images focusing on the In$_2$Se$_3$ layer of the sample are shown in Figures 1d and 2a again depicting a well-ordered highly crystalline material which seems free of any structural defects.

Previously, we found by using HR-XRD[19] that these In$_2$Se$_3$ layers had a rhombohedral crystal structure with the space group $R\bar{3}m$. It is well known that In$_2$Se$_3$ with this space group has two principal polymorphs, namely the α and β phases[15,14,22]. Since the lattice parameters of α-In$_2$Se$_3$ and β-In$_2$Se$_3$ are very similar, using HR-XRD, we were unable to determine the identity of the In$_2$Se$_3$ phase i.e., whether it was α, β or a mixture of the two. Attempts to determine the In$_2$Se$_3$ phase using Raman spectroscopy and photoluminescence spectroscopy also failed because the In$_2$Se$_3$ layer was too thin and the background signal from the InP wafer completely masked the signal from the layer.

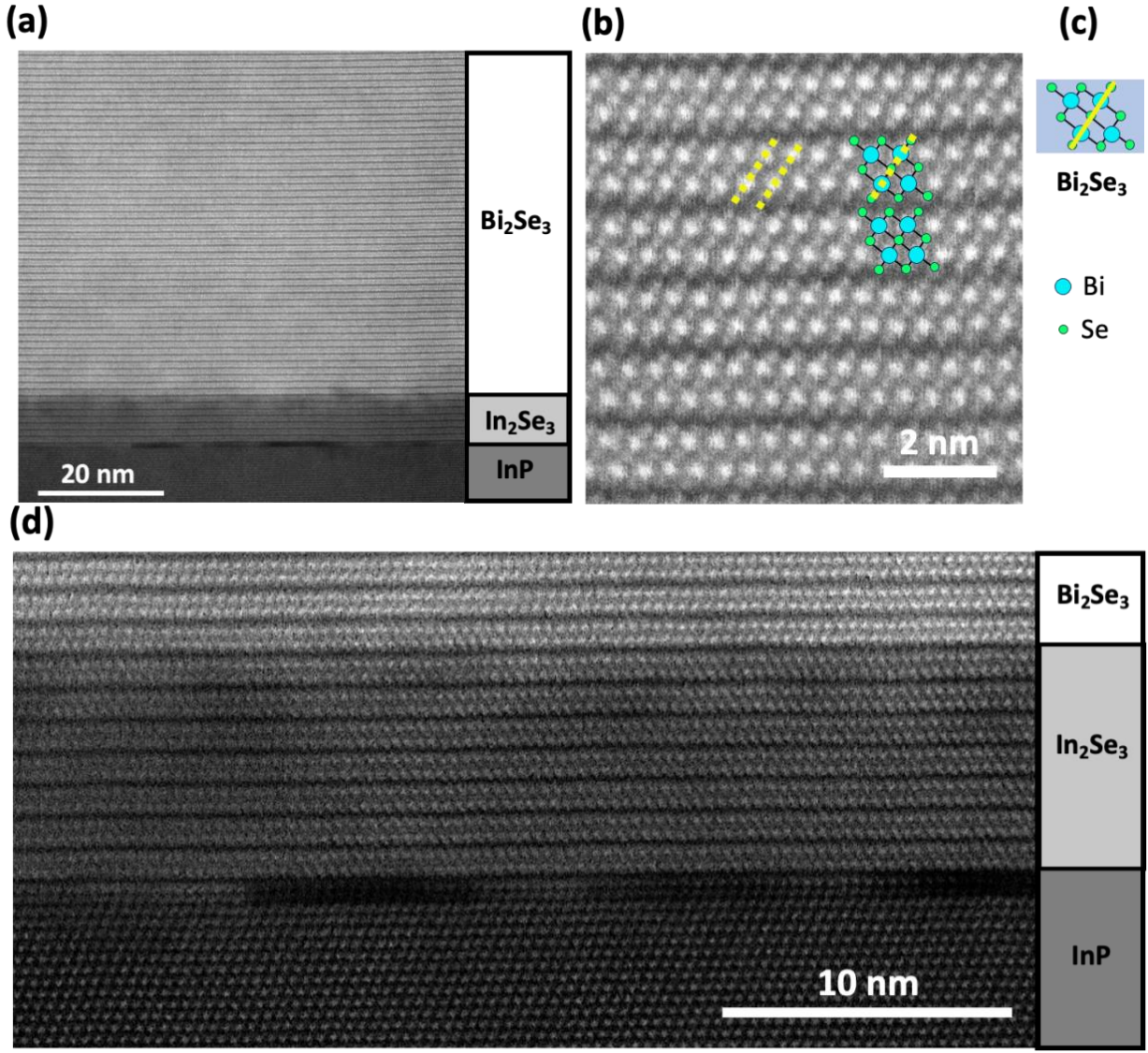

**Figure 1.** Cross-sectional HAADF-STEM images of Bi$_2$Se$_3$ (65-nm thick) grown on In$_2$Se$_3$/InP(111)B. (a) Highly ordered Bi$_2$Se$_3$ layer having an abrupt interface with the In$_2$Se$_3$ layer; (b) Enlarged image of Bi$_2$Se$_3$ layer with overlay of the atomistic model of Bi$_2$Se$_3$ quintuple layer. (c) Atomic model of B$_2$Se$_3$ quintuple layer consisting of 5 atoms. (d) Cross-sectional HAADF-STEM image showing well-ordered In$_2$Se$_3$ layer and interfaces with Bi$_2$Se$_3$ and InP.

Using cross-sectional HAADF-STEM images, as shown in Figure 2a, we established that the $In_2Se_3$ layer consists of a single phase. Furthermore, by using atomic-resolution HAADF-STEM images, such as shown in Figure 2b, and by overlaying atomistic models of α-$In_2Se_3$ and β-$In_2Se_3$ (see Figure 2c), we can conclude that the $In_2Se_3$ layer grown here is the β phase. The different phases of $In_2Se_3$ which can be formed readily, make it very difficult to achieve single phase material[23]. However, these results indicate that the $In_2Se_3$ layers grown using our newly developed technique[19] is capable of forming single-phase β-$In_2Se_3$.

Figure 2d shows an enlarged image of Figure 2a highlighting the interface between the $In_2Se_3$ and $Bi_2Se_3$ layers. In the image, we see evidence for indium diffusion into the $Bi_2Se_3$ layer as well as the formation of a single crystalline quintuple layer (1QL) with a mixture of both $In_2Se_3$ and $Bi_2Se_3$. In this intermixed layer, the 5-atom quintuple chain highlighted inside the red box is similar in size and shade to the $In_2Se_3$ region whereas those in the blue box have larger atoms at the In atomic column positions, which indicates these are Bi atoms that form $Bi_2Se_3$ quintuple chains in some regions. This intermixed layer is limited to 1QL and the $Bi_2Se_3$ layer above then makes an abrupt interface with this layer. We do not observe any interfacial layer of poor crystalline quality, neither diatomic steps nor $Bi_2Se_4$ clusters as seen at the interface that forms when $Bi_2Se_3$ is grown on flat InP or on rough non-vicinal InP substrates, as reported elsewhere[20,21].

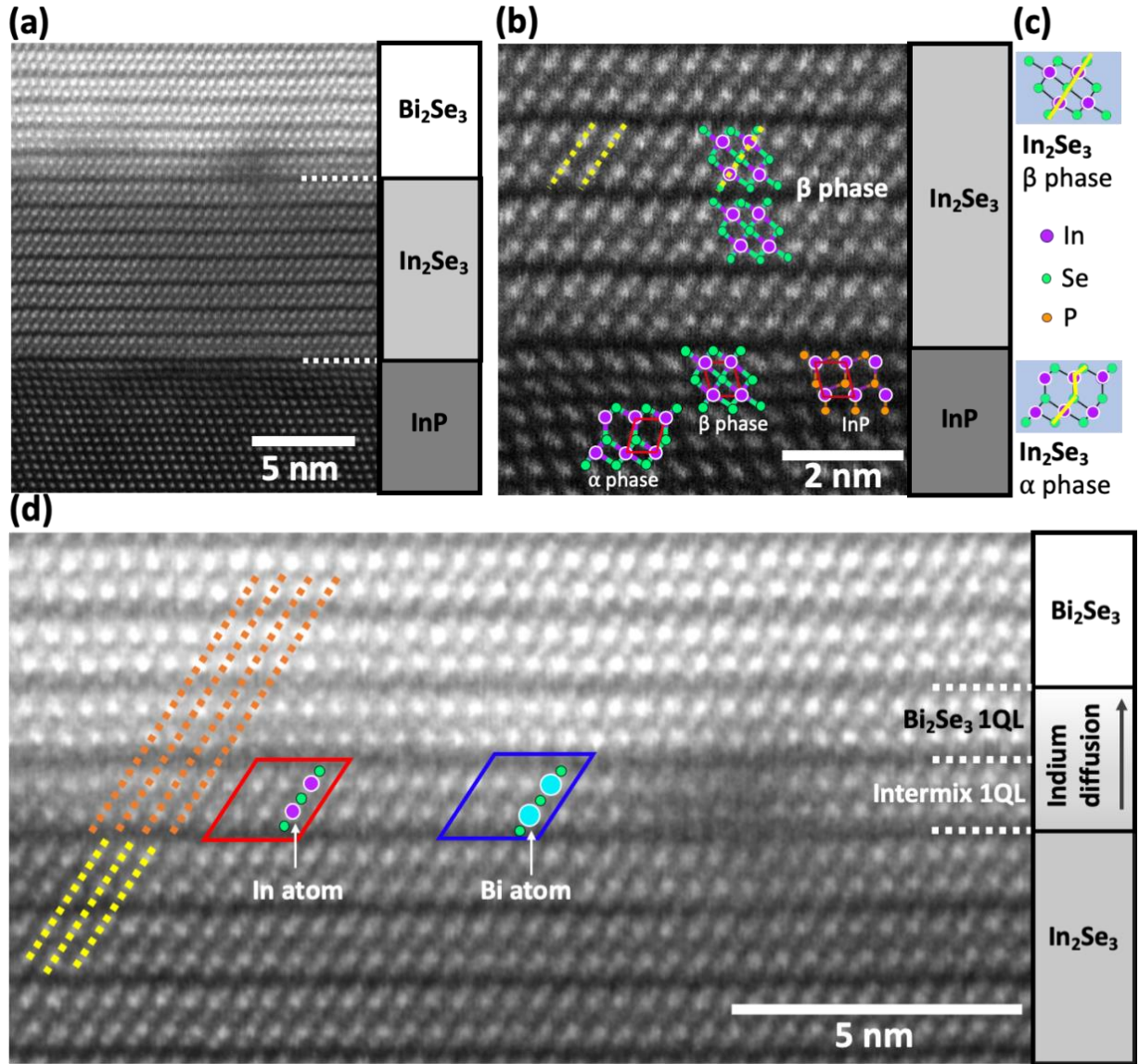

**Figure 2.** (a) Cross-sectional HAADF-STEM image of the sample taken at higher magnification than the image in Figure 1(d) showing the high crystallinity of β phase $In_2Se_3$ layer. (b) Enlarged image showing the atomic resolution of the β phase $In_2Se_3$ layer. An overlay of indium atoms in α and β phases of $In_2Se_3$ quintuple layer with indium atoms in InP lattice are also shown. Atomic model of InP(111)B lattice is shown next to the β-$In_2Se_3$ atomic model. (c) Atomic model of α and β phases of $In_2Se_3$ quintuple layer which consists of 5 atoms. (d) Enlarged image of Figure 2(a) showing In diffusion into $Bi_2Se_3$ layer at $Bi_2Se_3$/$In_2Se_3$ interface. The quintuple layer highlighted in the red box shows 5-atom chain $In_2Se_3$ whereas 5-atom chain $Bi_2Se_3$ is shown in the blue box. Orange and yellow lines are for guiding the eyes.

The interface between the In$_2$Se$_3$ layer and the InP substrate has several notable features. As visible in the HAADF-STEM image shown in Figure 1a and the higher magnification image, Figure 1d, the interface has dark regions alongside the clear and sharp interface, suggesting the presence of some possibly defective regions. A significant contrast feature visible in some higher magnification cross-sectional bright field (BF) STEM images (Figures 3a and b) of the sample suggests significant Se diffusion into the substrate beyond the In$_2$Se$_3$/InP interface. The presence of excess Se, as evident from the contrast difference near the interface, does not however alter the crystal structure of the InP substrate, which remains zinc blende. This observation suggests that the In atoms remain fixed in their lattice sites during the In$_2$Se$_3$ formation process while Se diffuses into the substrate, displacing P. Then, with the special annealing step during the growth process, the Se atoms re-arrange to form In$_2$Se$_3$ quintuple layers. Figure 3c is a schematic illustrating the In$_2$Se$_3$ formation steps during oxide desorption from the InP(111)B substrate surface in the unique MBE-based growth approach used here[19].

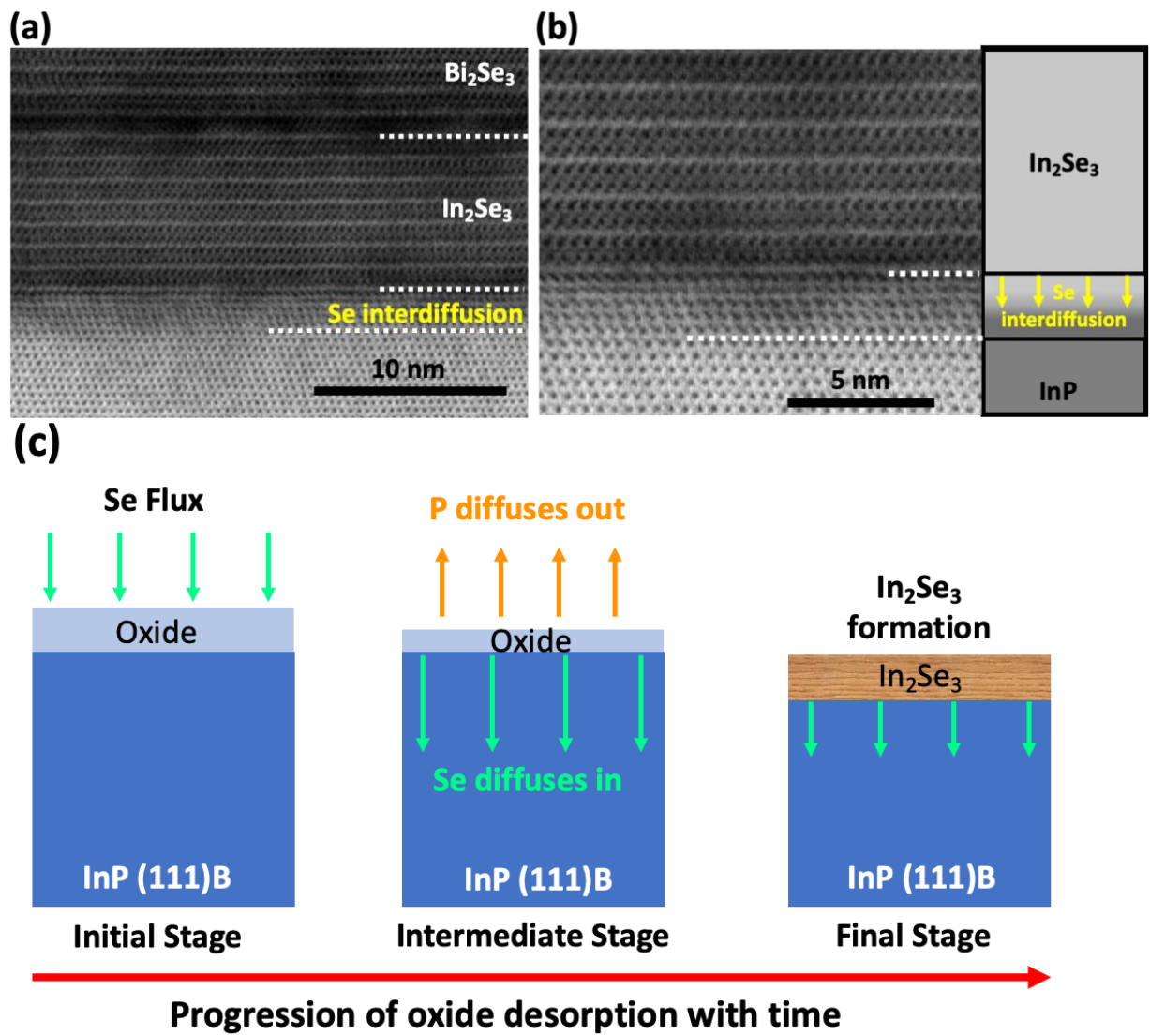

**Figure 3.** (a) Cross-sectional BF-STEM images highlighting diffusion of Se atoms into InP substrate during the $In_2Se_3$ growth process. (b) Cross-sectional BF-STEM image at higher magnification showing that the presence of excess Se atoms does not alter the crystal structure of the InP, which remains zinc blende. (c) Simple schematic showing Se atoms diffuse into InP substrate while P atoms diffuse out during the oxide desorption process and the formation of $In_2Se_3$ as the oxide desorption process is completed.

If this proposed mechanism, in which In atoms remain "anchored" through the process, is valid, then the preference to form the single β-In$_2$Se$_3$ phase over the α-In$_2$Se$_3$ phase can be understood. In Figure 2b, we have overlayed the In atoms in the α and β phases of In$_2$Se$_3$ quintuple layers on top of the In atoms in the InP(111)B lattice. We observe that the 4 In atoms in the red oblique square of the β-In$_2$Se$_3$ are well superimposed on top of the In atoms in the InP lattice whereas only the top two In atoms out of the four In atoms are superimposed in the red oblique square of the α-In$_2$Se$_3$, while the bottom two are shifted away from the In atoms in the InP lattice. This careful observation also leads to the conclusion that the In atoms remain anchored during the In$_2$Se$_3$ formation process Based on these observations of cross-section STEM images as discussed above, the proposed growth mechanism for the formation of the twin-free and single-phase β-In$_2$Se$_3$ is illustrated using a cartoon atomic model in Figure 4 where the transformation of the 3D InP lattice into 2D β-In$_2$Se$_3$, via an intermediate Se substituted zinc blende structure, is apparent in this growth process.

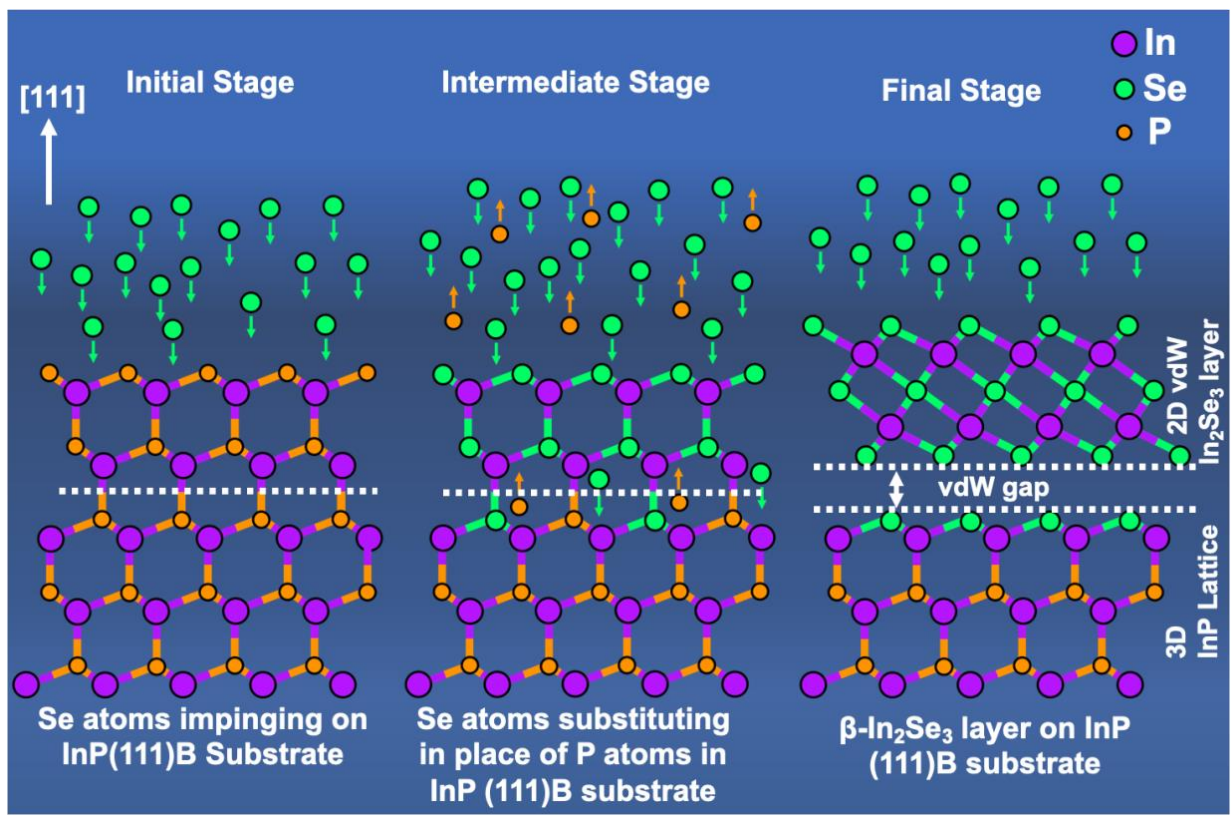

**Figure 4.** Proposed growth mechanism demonstrated using a cartoon atomic model for the formation of untwinned single-phase In$_2$Se$_3$ layer based on the STEM image analysis. Orange arrows and green arrows represent desorption of P atoms from the substrate, and diffusion of Se atoms into the substrate, respectively.

In Summary, we have shown that high quality In$_2$Se$_3$ and Bi$_2$Se$_3$ crystalline layers that are fully twin-free and largely free of defects, can be achieved. STEM observations also show that the sample consists primarily of single phase β-In$_2$Se$_3$. Close observation of the InP/In$_2$Se$_3$ interface provides evidence for an exchange mechanism in which Se first substitutes for P in the zinc blende InP, followed by a crystal structure transformation to the rhombohedral In$_2$Se$_3$ structure. This result implies that In atoms are not mobile during the transformation, thus resulting in twin-free In$_2$Se$_3$. This unique growth mechanism for In$_2$Se$_3$ also results in pure β-phase In$_2$Se$_3$. Controlling the mechanism at play would enable the application of this approach to other heteroepitaxial structures involving layered, vdW materials on 3D crystalline substrates. It may also enable the possibility of stabilizing the different phases of $In_2Se_3$ which are of important technological interest owing to their exotic properties.

## DATA AVAILABILITY

All data are available from the corresponding author upon reasonable request.

## CORRESPONDING AUTHOR

Correspondence and request for materials should be addressed to M.C.T

## AUTHOR CONTRIBUTIONS

K.S.W. conceived and executed the research and MBE growth. K.S.W. and M.C.T. wrote and reviewed the manuscript. M.R.M and D.J.S were responsible for the relevant sample preparation and STEM imaging and reviewed the manuscript. All authors contributed to interpretation of the data and discussions.

## NOTES

The authors declare no competing interests.

## ACKNOWLEDGMENTS

This work was supported by NSF grant number HRD-2112550 (NSF CREST Center IDEALS). Partial support is also acknowledged from NSF grant number DMR-2011738 (PAQM). The authors acknowledge use of facilities in the John M. Cowley Center for High Resolution Electron Microscopy at Arizona State University, supported in part by NNCI-ECCS-1542160.

# Electronic Supplementary Information

## 1. Materials and Instrumentation

Samples were grown on smooth, non-vicinal Fe-doped InP(111)B ± 0.5° substrates with a phosphorus-terminated surface. A Riber 2300P molecular beam epitaxy (MBE) system with a base pressure of $5\times10^{11}$ Torr was used to grow the material. The system was equipped with *in situ* reflection high-energy electron diffraction (RHEED) to facilitate monitoring of material growth in real time. High-purity (99.9999%) bismuth (Bi) and selenium (Se) were used as source materials. The Bi and Se fluxes were provided by a RIBER dual-zone effusion cell and a RIBER valved cracker cell for corrosive materiel (VCOR), respectively. An ultra-high vacuum (UHV) nude ion gauge positioned to intercept the path of the fluxes was used to measure beam equivalent pressures (BEP).

The samples were characterized using high-angle annular-dark-field (HAADF) and bright-field (BF) scanning transmission electron microscopy (STEM) imaging using a probe-corrected JEOL ARM200F operated at 200kV. Cross-section samples suitable for TEM observation were prepared using a dual-beam Thermo Fisher Helios 5G UX gallium ion milling system, initially at 30keV with subsequent thinning at 5keV and 2keV to reduce the amount of surface damage.

## 2. Synthesis

A $Bi_2Se_3$ layer was grown on an $In_2Se_3$ layer, which was formed using a unique and non-conventional Se self-passivation technique developed in our group[1]. The $In_2Se_3$ layer was formed at a substrate temperature of $T_{sub} = 505$ °C with an Se overpressure of $1\times10^{-5}$ Torr during the oxide desorption process without an indium (In) source cell. The $Bi_2Se_3$ growth temperature ($T_{sub}$), growth rate and BEP ratio of Se to Bi were $T_{sub} = 270$ °C, ~0.5 nm/minute and ~100:1 respectively. A 1x1 RHEED pattern was observed after $Bi_2Se_3$ growth with increased intensity relative to the $In_2Se_3$ layer. The newly developed growth procedure for the formation of the $In_2Se_3$ layer and the twin-free nature of $In_2Se_{3\bar{5}}$ as well as $Bi_2Se_3$, were previously discussed in detail by Wickramasinghe et al[1]. The substrate temperature was measured using a thermocouple mounted behind the substrate holder. The sample analyzed here consisted of a 65-nm-thick $Bi_2Se_3$ layer on top of a ~ 7-nm-thick $In_2Se_3$ layer, both of which are twin-free[1].

## 3. HAADF images

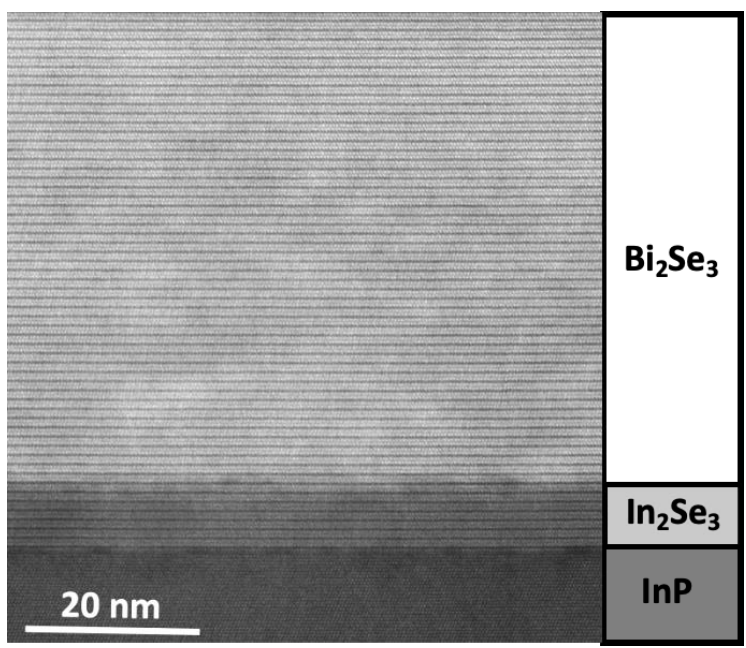

**Supplemental Figure 1.** Cross-sectional HAADF-STEM images of $Bi_2Se_3$ (65-nm thick) grown on $In_2Se_3$/InP(111)B taken from a region several microns away from the area shown in Figure 1.a.

## References

1. Wickramasinghe, K. S., Forrester, C. & Tamargo, M. C. Molecular Beam Epitaxy of Twin-Free $Bi_2Se_3$ and $Sb_2Te_3$ on $In_2Se_3$/InP(111)B Virtual Substrates. *Crystals* **13**, (2023).